\documentclass[5p,sort&compress]{elsarticle}
\usepackage{amsmath}
\usepackage[colorlinks = true, linkcolor = blue, citecolor = blue]{hyperref}
\usepackage[mathlines]{lineno}
\usepackage{graphicx}
\usepackage{dblfloatfix} 
\usepackage{lipsum}
\usepackage{color}
\usepackage{soul}
\usepackage[table]{xcolor}
\usepackage{tabularx} % for textwidth tables
\usepackage{tikz}
\usepackage{float}
\usepackage{array}
\usepackage{booktabs}
\modulolinenumbers[5]
\usepackage[numbers]{natbib}
\hypersetup{       
	unicode=false,          
	pdftoolbar=true,        
	pdfmenubar=true,        
	pdffitwindow=true,     
	pdfstartview={FitH},    
	pdftitle={Design and optimization of a 100 keV DC/RF ultracold electron source}, 
	pdfauthor={D.F.J. Nijhof},     
	pdfsubject={},   
	pdfcreator={D.F.J. Nijhof},   
	pdfproducer={D.F.J. Nijhof}, 
	pdfkeywords={}, 
	pdfnewwindow=true,      
	colorlinks=true,       
	linkcolor=blue,          
	citecolor=purple,        
	filecolor=cyan,      
	urlcolor=magenta           
}

\journal{Nucl. Instrum. Methods Phys. Res. A}
\begin{document}
	\begin{frontmatter}
		
%		\title{Design and optimization of a 100 keV DC/RF pulsed, ultracold, and ultrafast electron source}
		\title{Design and optimization of a 100 keV DC/RF ultracold electron source}
		\cortext[mycorrespondingauthor]{Corresponding author}
		\author[1]{D.F.J. Nijhof\corref{mycorrespondingauthor}}
		\ead{d.f.j.nijhof@tue.nl}
		\author[1]{P.H.A. Mutsaers}
		\author[1,2,3]{and O.J. Luiten}

		\address[1]{Department of Applied Physics and Science Education, Coherence and Quantum Technology Group, Eindhoven University of Technology, P.O. Box 513, 5600 MB Eindhoven, the Netherlands}
		\address[2]{Institute for Complex Molecular Systems, Eindhoven University of Technology, P.O. Box 513, 5600 MB Eindhoven, The Netherlands}
		\address[3]{Doctor X Works BV, 5616 JC Eindhoven, The Netherlands}
		
		\begin{abstract}
			An ultracold electron source based on near-threshold photoionization of a laser-cooled and trapped atomic gas is presented in this work. Initial DC acceleration to $\sim$10 keV and subsequent acceleration of the created bunches to 100 keV by RF fields makes the design suitable to serve as injector for accelerator-based light sources, single-shot ultrafast protein crystallography, applications in dielectric laser acceleration schemes, and potentially as an injector for free electron lasers operating in the quantum regime. This paper presents the design and properties of the developed DC/RF structure. It is shown that operation at a repetition frequency of 1 kHz is achievable and detailed particle tracking simulations are presented showing the possibility of achieving a brightness that can exceed conventional RF photosources.
		\end{abstract}

		
	\end{frontmatter}
%	\linenumbers
\section{INTRODUCTION}\label{sec:introduction}

\begin{figure*}[h!]
	\centering
	\includegraphics[width=1\textwidth]{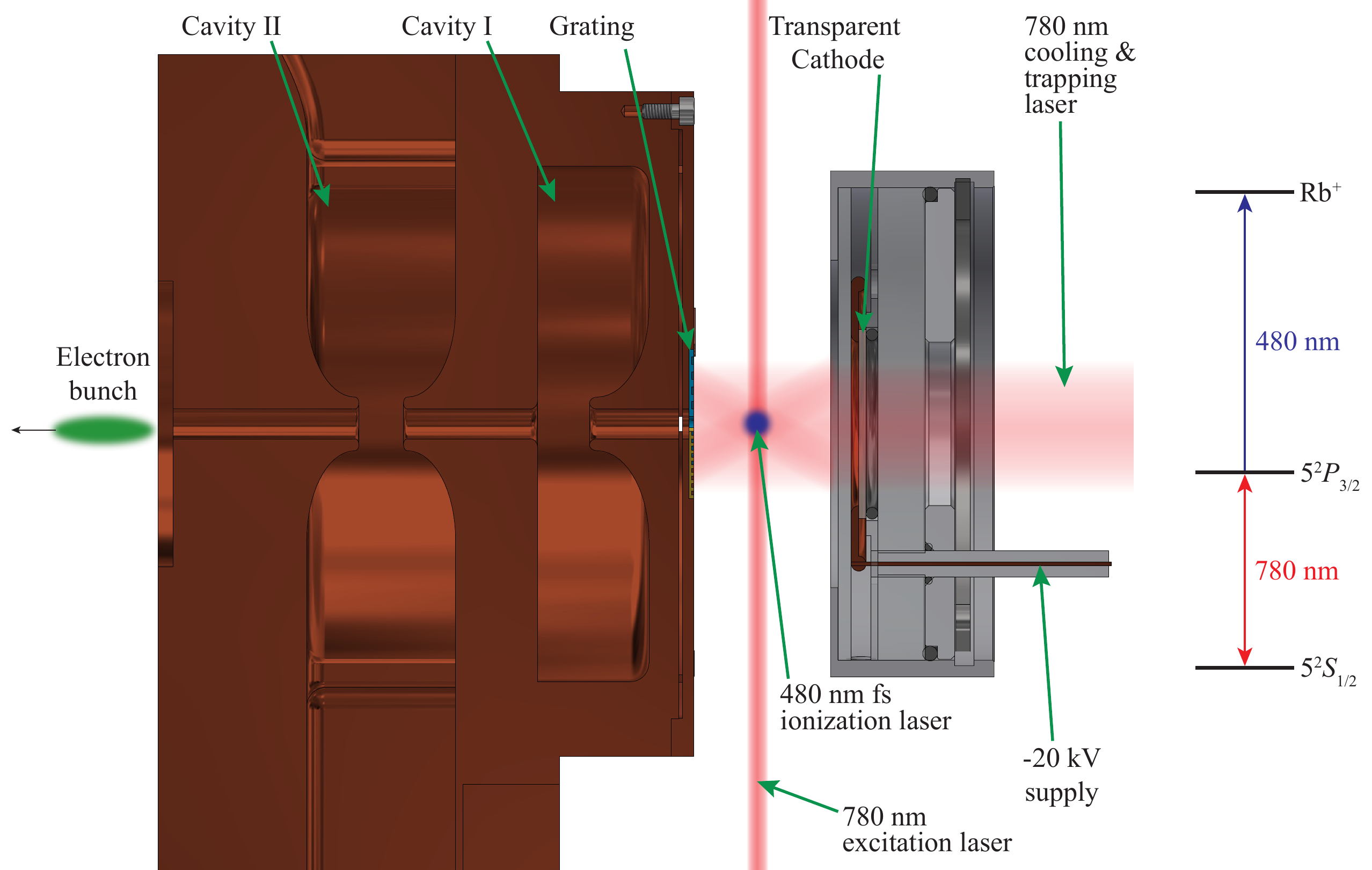}
	\caption{Schematic representation of the proposed source showing from left to right: the RF structure, the ionization region with cooling, trapping, and ionization lasers, the accelerator module, and the excitation-ionization scheme.}
	\label{fig:schematic_1_2}
\end{figure*}
%%%%%%%%%%%%%%%%%%%%%%%%%%%%%%%%%%%%%%%%%%%%%%%%%%%%%%%%%%%%%%%%%%%%%%%%%%%%%%%%%%%%%%%%%%%%%%%%%%%
\indent In recent years an ultrafast and ultracold electron source (UCES) based on two-step, near-threshold photoionization of a laser-cooled and trapped atomic gas has been developed \cite{Claessens2005, Taban_2010, McCulloch2011, Engelen_2013, McCulloch2013, Engelen_2014, Speirs2015a, Franssen_2019_2}. The goal of a source of this type has been to offer a combination of a high beam quality and high bunch charge.
% whilst maintaining a large source size. 
Sub-ps electron bunches with an energy of 10 keV and normalized transverse emittances in the order of a few nm rad have been measured \citep{Franssen_2019_2}. Building upon this idea, a source is proposed that combines the benefits provided by the UCES with a radio frequency (RF) structure that increases the bunch energy to $\sim$100 keV whilst maintaining the excellent beam quality.\\
\indent Applications for the proposed source are found in fields such as ultrafast electron diffraction (UED) where ultrafast dynamics of matter at the atomic scale can be studied through pump-probe experiments. A high quality electron bunch in terms of the transverse coherence length is required for such applications \cite{Zewail_2010}. To realize this for a given beam size, an ultracold electron source is needed. The proposed source is capable of producing $>$ 10 fC bunches with a transverse normalized emittance in the order of $\sim$nm rad such that even (single-shot) protein crystallography could be realized.\\
\indent Furthermore, the design presented in this work may be considered as a suitable injector for an FEL operating in the quantum regime \cite{Schaap_2022}, or in the field of dielectric laser acceleration (DLA) where typically low-charge bunches are injected in microscopic structures, where very high transverse quality bunches are required \cite{Hommelhoff_2022}.\\
\indent The main selling point of the proposed source is based on the low transverse emittance of the electron bunches extracted from a laser-cooled atomic gas \cite{Franssen_2019_2}. The bunches are however extracted at an energy of only $\sim$10 keV. Immediate acceleration after extraction should prove beneficial for bunch quality preservation as it reduces beam degradation due to Coulomb interactions.\\
\indent For the structure presented in this work an emphasis is put on the design being simple, compact, and robust. Having only a few accelerating cells greatly simplifies both the manufacturing process and reduces the cost. Using a single waveguide to supply RF power (which will henceforth be referred to as the 'RF feed') ensures phase stability and magnetic coupling through the wall connecting the two cells guarantees phase synchronization between the cells. Powering the structure and accelerating electrons from 10 keV to 100 keV requires only a 5 kW solid-state RF amplifier.\\
\indent This paper presents the design, optimization, EM field simulations, and particle tracking simulations of a hybrid DC/RF structure \cite{Geer_2014}, consisting of a grating magneto-optical trap (MOT) in a static extractor field \cite{Nshii_2013, Franssen2017} and a 1$\frac{3}{4}$-cell standing wave RF cavity. Source requirements and a general overview of the source are presented in Sec. (\ref{sec:source_requirements}), the source design and optimization strategy is presented in Sec. (\ref{sec:optimization}) along with electromagnetic field simulations and a thermal analysis. Finally, realistic particle tracking simulations are presented in Sec. (\ref{sec:particle_tracking_simulations}).\\
\section{SOURCE REQUIREMENTS AND OVERALL DESIGN}\label{sec:source_requirements}

The electron source presented in this paper builds on an existing DC electron source, as well as other similar setups \cite{Claessens2005, Taban_2010, McCulloch2011, Engelen_2013, McCulloch2013, Engelen_2014, Speirs2015a, Franssen_2019_2} and a proposal for a hybrid DC/RF electron source \cite{Geer_2014}. The applications mentioned in Sec. (\ref{sec:introduction}) all require a high bunch quality and common for all applications is the requirement for a low transverse normalized emittance, which is given by the following:
\begin{equation}\label{eq:normalized_emittance}	
	\varepsilon_{nx}=\frac{1}{mc}\sqrt{\left<x^{2}\right>\left<p_{x}^{2}\right>-\left<xp_{x}\right>^{2}},
\end{equation}
where $\left<...\right>$ denotes the averaging of a parameter over the electrons in the bunch, $x$ the transverse position, $p_{x}=mc\gamma\beta_{x}$, with $m$ the electron rest mass, $c$ the speed of light, $\gamma$ the Lorentz factor, and $\beta_{x}=v_{x}/c$ the normalized electron bunch's velocity in the $x$-direction with respect to the speed of light.\\
\indent The structure presented in this paper is designed to accelerate electron bunches to an energy of 100 keV, a commercially available solid-state RF amplifier (5 kW peak power) should provide the structure with sufficient power to accelerate the bunches to the desired energy, enabling repetition rates of $\geq1$ kHz.\\
\indent A schematic representation of the proposed source is shown in Fig. (\ref{fig:schematic_1_2}) with, from left to right: the copper RF structure housing the two cavities and a mounted grating chip, the region in which laser cooling and trapping occurs, and the accelerator module housing the -20 kV cathode. Electron bunches are created in the region between the cathode and the RF structure where the excitation and ionization lasers overlap.

\begin{figure*}[h!]
	\centering
	\includegraphics[width=0.8\textwidth]{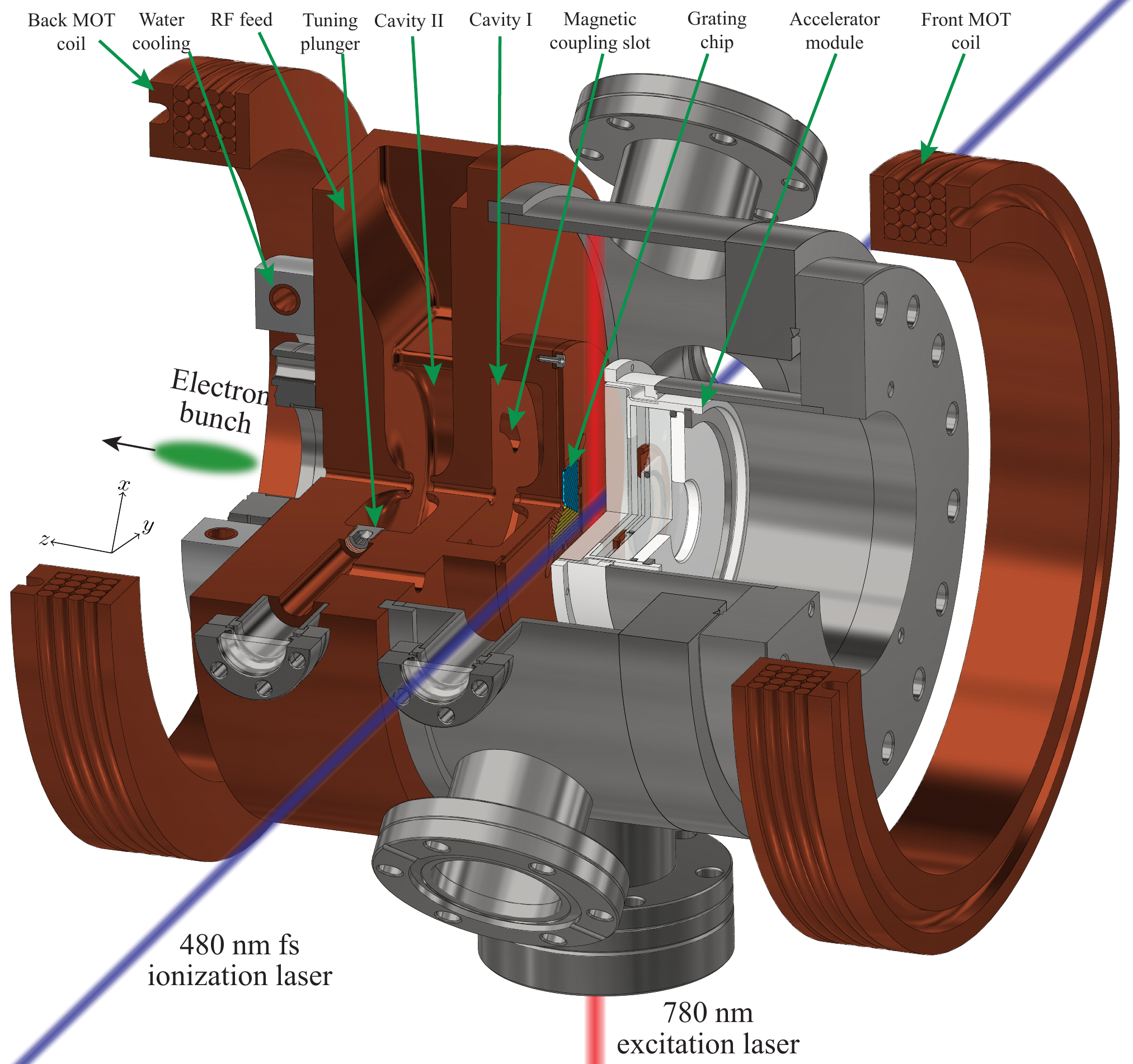}
	\caption{Impression of the proposed setup showing the main components of the electron source (cooling and trapping laser not shown).}
	\label{fig:schematic_3_4}
\end{figure*}
A MOT is created by a single trapping laser beam, which passes through the ITO electrode (transparent cathode). This trapping laser at $\lambda\sim780$ nm is diffracted on a grating chip consisting of three linear gratings oriented 120 degrees with respect to each other \cite{Franssen_2019_2, Nshii_2013}. The diffracted first orders create, together with the incident laser beam, a diamond-shaped region where laser-cooling of the $^{85}$Rb atomic gas can be realized (in combination with an anti-Helmholtz magnetic coil configuration (see Fig. (\ref{fig:schematic_3_4})).
\\
\indent The atoms are ionized in a two-step process, shown in Fig. (\ref{fig:schematic_1_2}) on the right. They are first excited by the 780 nm excitation laser (red), which pumps the $5^{2}S_{1/2}F=3\rightarrow5^{2}P_{3/2}F=4$ transition, and the femtosecond (fs) 480 nm ionization laser (blue) which ionizes the laser-cooled and trapped $^{85}$Rb gas, resulting in a ps duration electron bunch at an energy of $\sim$10 keV \citep{Raadt_2023}. This bunch is then accelerated to 100 keV by the RF part of the source. The RF part is a 1$\frac{3}{4}$-cell, magnetically coupled (not shown in this figure), standing wave, S-band (2.99855 GHz), $\pi$-mode accelerator structure.\\
\indent The two cells in the RF structure are magnetically coupled. Coupling through the cavity walls is necessary to minimize the required amount of RF power. Nose-cones in the cells ensure a sufficiently large shunt impedance. Power is fed into the second cell through a coupling slot with a modified WR284 waveguide. An auxiliary waveguide with a similar coupling structure located underneath the feed waveguide significantly reduces any induced dipole fields and also functions as a port for a vacuum pump. A tuning plunger is located in the second acceleration cell, which allows for tuning of the resonant frequency of the structure by $\pm$2 MHz. An impression of the source, complete with MOT coils, viewports for the lasers, and water cooling is shown in Fig. (\ref{fig:schematic_3_4}).

\section{ACCELERATOR STRUCTURE}\label{sec:optimization}
This section will discuss the accelerating structure used in the proposed source and describe the design and optimization of the newly designed RF structure. First, the DC-based UCES is briefly introduced and its merits discussed in section (\ref{subsec:DC_structure}), followed by the newly designed RF structure, where the geometry, power feed, electromagnetic field distribution, and thermal properties of the source are presented in section (\ref{subsec:RF_structure} - \ref{subsec:thermal_analysis}) respectively.

\subsection{DC structure}\label{subsec:DC_structure}

As mentioned earlier, the DC part of the proposed hybrid DC \& RF electron source is based on a readily existing source \cite{Franssen_2019_2}. This source produces electron bunches with a temporal length in the order of picoseconds at electron temperatures of $\sim$ 10 K \cite{Franssen2017, Raadt_2023}. It does so through a two-step photoionization process with a broadband fs laser pulse ($\sim40$ nm FWHM). Being able to scan the central wavelength of this broadband spectrum enables the production of low excess energy electrons. The laser cooling takes place in a grating based magneto optical trap \cite{Nshii_2013}. These bunches are created in the center of a static field with a potential of -20 kV applied across a distance of 19.2 mm, resulting in an accelerating gradient of 1.4 MV/m \cite{Franssen_2019_2}.\\
\indent Creating electrons at higher energies is possible by increasing this acceleration voltage. There is however a practical limit for this due to dissimilar Stark shifts of the hyperfine levels of the excited $5^{2}P_{3/2}$ state, which can be avoided by keeping the acceleration potential at $<2$ MV/m \cite{Franssen_2019, Krenn_1997}. Therefore the applied potential is limited to -20 kV and additional acceleration is realized through an RF accelerator. The attainable repetition frequency of this source is not limited by the heating of a photo cathode but by the fs laser repetition frequency and the filling time of the MOT from which the electron bunches are extracted, allowing operation at $\geq1$ kHz

\subsection{RF structure}\label{subsec:RF_structure}
\begin{figure}[t!]
	\includegraphics[width=1\linewidth]{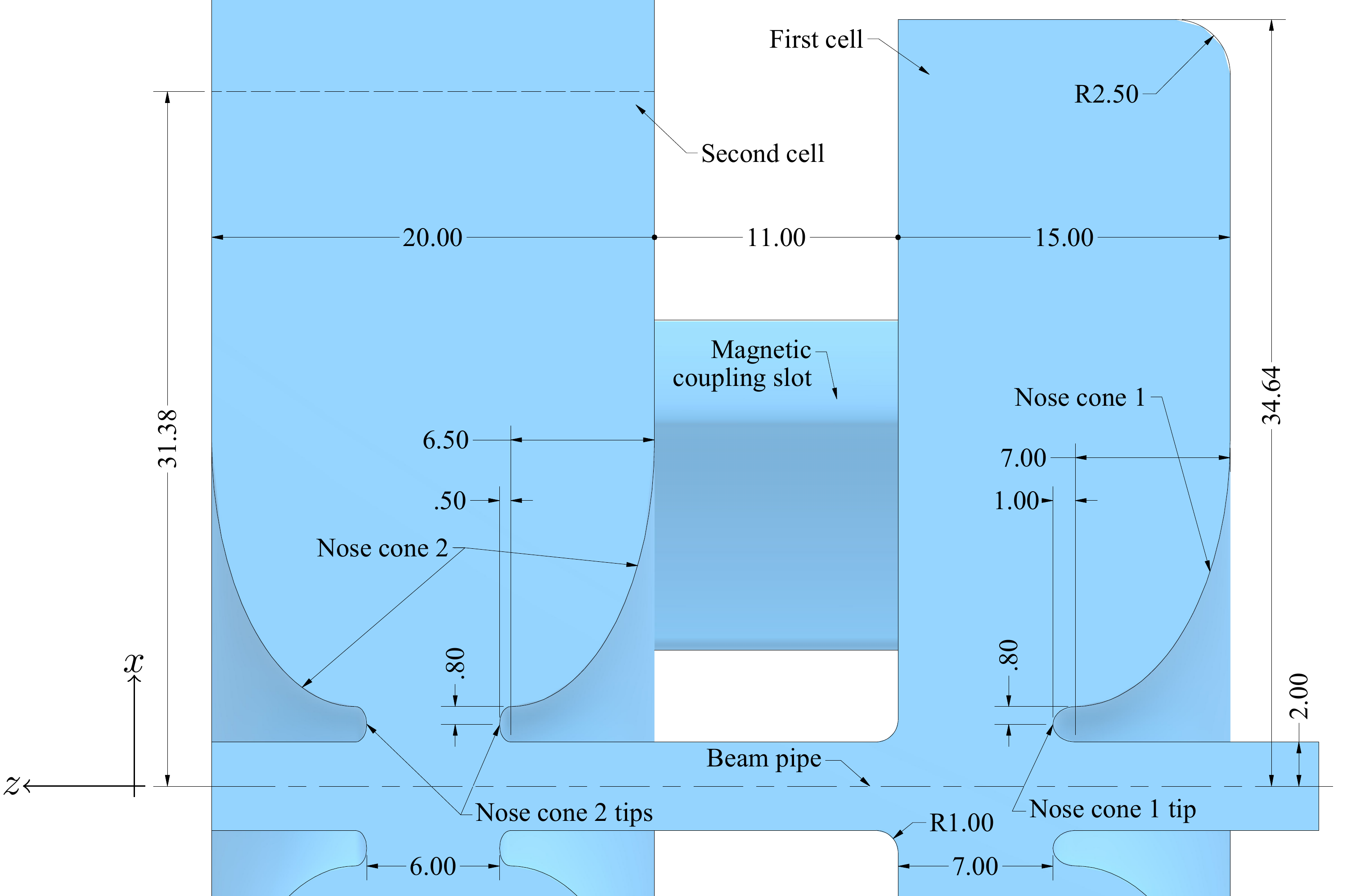}
	\caption{Schematic representation of the (vacuum) geometry of the RF structure in the $x$-$z$ plane (all dimensions are in millimeters).}
	\label{fig:cavity_dimensions_x_z}
\end{figure}
\begin{figure}[b!]
	\includegraphics[width=1\linewidth]{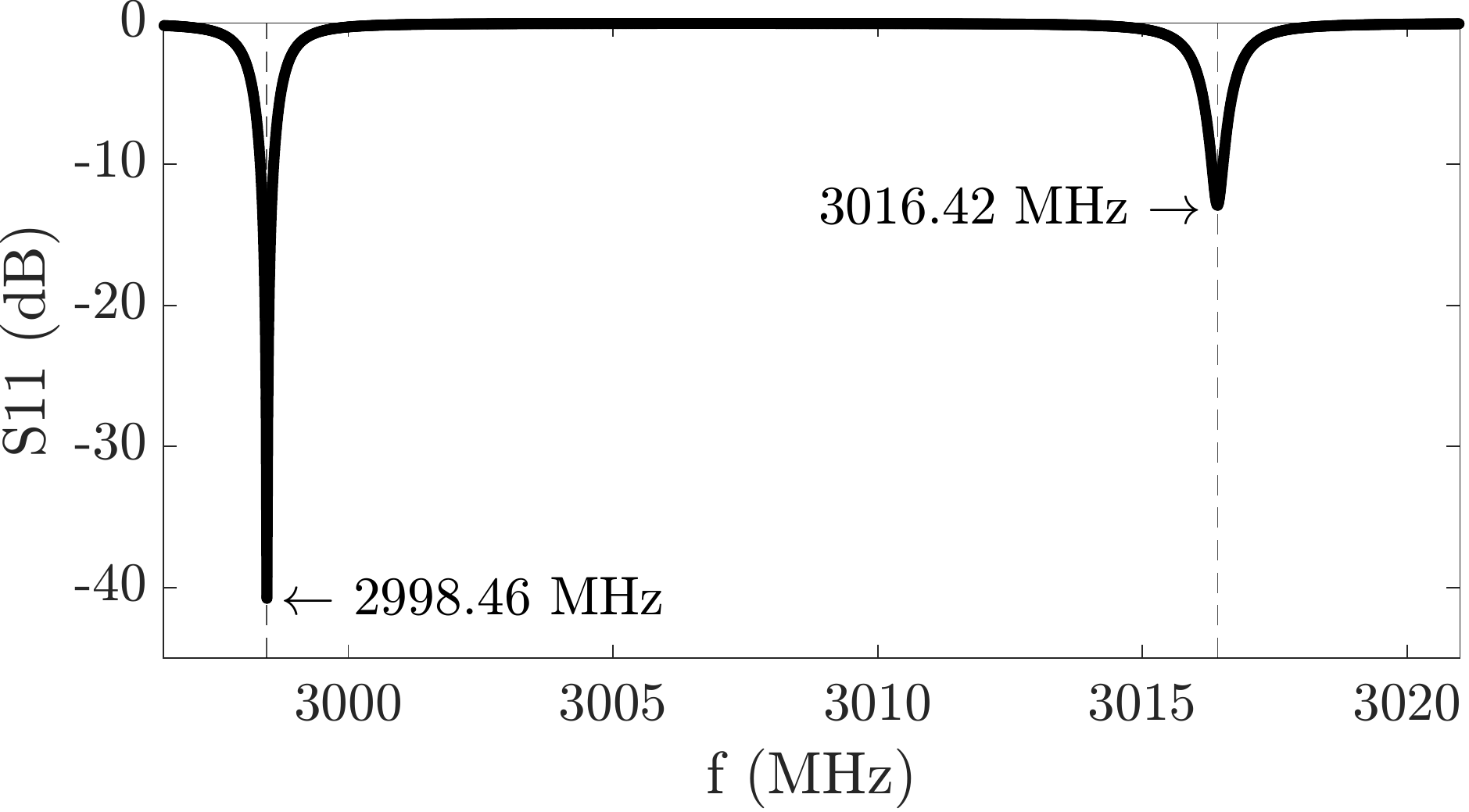}
	\caption{The absorption plot of the RF structure with the two resonances at 2998.46 MHz (the desired $\pi$-mode) and 3016.42 MHz (the zero-mode) clearly visible.}
	\label{fig:S11_double_absorption}
\end{figure}
\begin{figure}[t!]
	\includegraphics[width=1\linewidth]{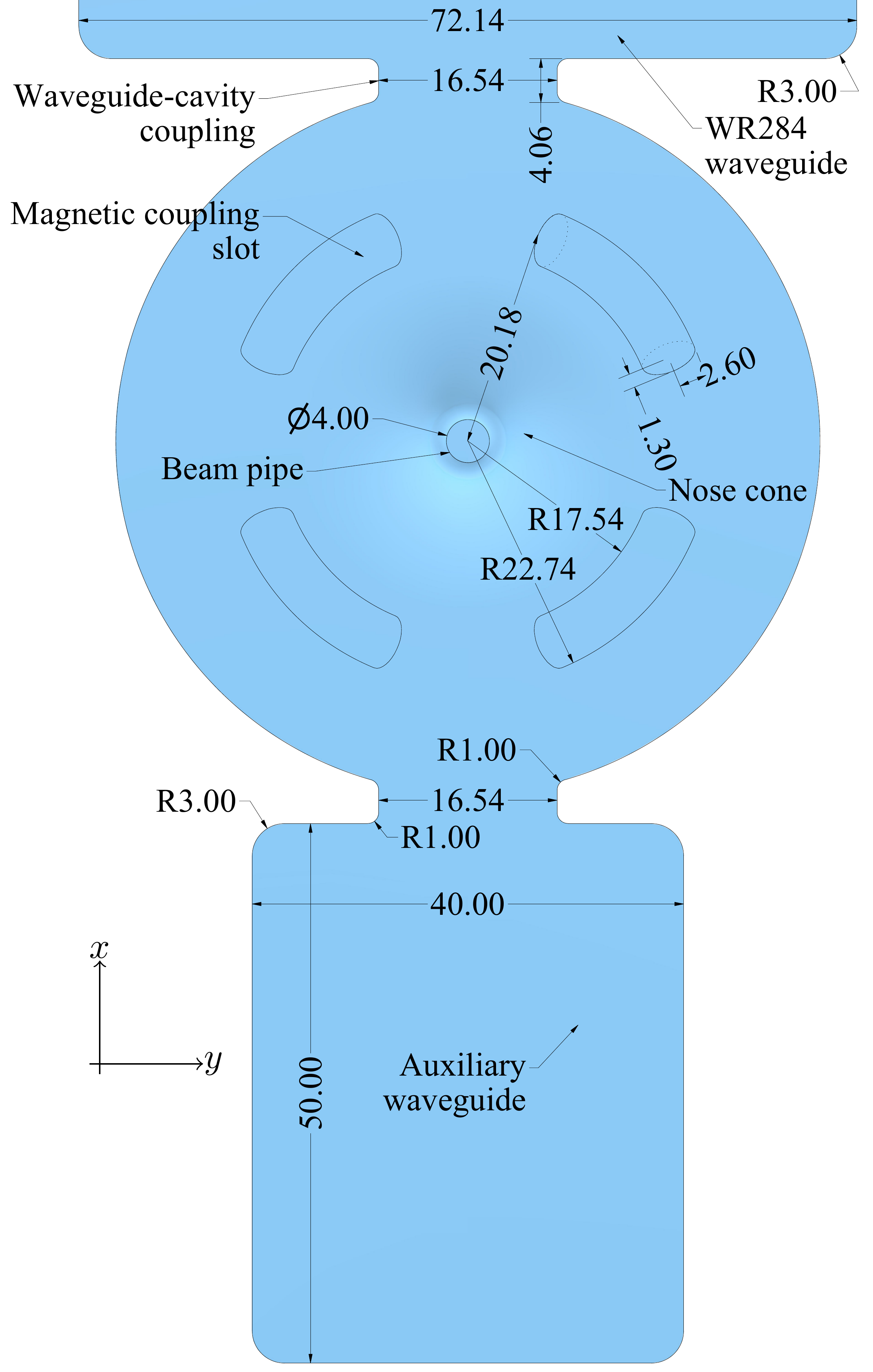}
	\caption{Schematic representation of the second cell's geometry and its coupling to the RF port in the $x$-$y$ plane (all dimensions are in millimeters).}
	\label{fig:cavity_dimensions_x_y}
\end{figure}

A schematic representation of the RF structure's geometry in the $x-z$ plane is shown in Fig. (\ref{fig:cavity_dimensions_x_z}). A multi-cell structure will always have a number of resonant frequencies equal to the amount of cells. These resonance frequencies will be visible in its absorption spectrum and are shown in Fig. (\ref{fig:S11_double_absorption}), obtained from the frequency domain solver of \textsc{cst microwave studio} \cite{CST}. 
The absorption peak at $f_{\pi}=2998.46$ MHz is the desired $\pi$-mode of the structure with an input reflection of approximately -40.8 dB. The undesired zero-mode is found at $f_{0}=3016.42$ MHz, giving a mode separation of roughly 18 MHz.\\
\indent The structure shown in Fig. (\ref{fig:cavity_dimensions_x_z}) is initially optimized by varying the major axes of the nose cones in order to minimize the power dissipation in the copper, increasing the quality factor $Q$ of the structure. The width of the cell acceleration gaps (distance between the nosecone and cavity wall for the first cell and the distance between both nose cones for the second cell) is fixed at 7 mm and 6 mm, respectively.\\
\indent The power dissipation is minimized by varying the minor and major axes of the elliptical nose cone tips. The nose cone structures in both cells are necessary because of the low velocity of the electron bunches (the electrons enter the RF field with $\beta_{z}\approx0.195$). The nose cones reduce the length along which the electrons are accelerated, ensuring a sufficiently short transit-time. Additionally, the nose cones increase the shunt impedance, utilizing the available power more efficiently.\\
\indent The nose cones and small pipe radius prohibit coupling along the optical axis of the cavity so magnetic coupling of the cells through the shared side is necessary. These coupling slots are centered at the radial position where the magnetic fields have a large field strength in the TM$_{010}$-like mode of operation (see Sec. (\ref{subsec:fields})). The width of the coupling slots (distance between two two cavities) is of importance for the synchronization of the electron-field interaction between the cavities. The resonance frequency of the structure is set by varying the radii of both cells simultaneously, after which the desired ratio between the on-axis acceleration field strengths is set by varying the radius of the first cell.\\
\indent Optimizing the coupling of the RF feed (reduced WR284 waveguide: 72.136 mm by 20 mm) is critical since only a 5 kW (peak) pulsed solid-state amplifier is intended to be used. An $x$-$y$ projection of the second cell, the magnetic coupling slots, and waveguide feed is shown in Fig. (\ref{fig:cavity_dimensions_x_y}). It also features a coupling structure at the bottom of the cavity, thus minimizing any dipole fields induced by the coupling structure of the RF feed \cite{Chae_2011}. Optimizing the coupling between the top waveguide and the second cell is done primarily by varying the width of the waveguide coupler (16.54 mm in Fig. (\ref{fig:cavity_dimensions_x_y})), which is the same for both the bottom and top coupler structure. RF power at a frequency of $\sim$3 GHz can propagate freely through the top waveguide with a width of 72.136 mm, the bottom auxiliary waveguide width is smaller, resulting in a higher cut-off frequency, preventing it from absorbing RF power.\\
\begin{table}[b!]
	\caption{Overview of the operational parameters of the proposed DC/RF source.}
	\begin{tabularx}{\linewidth}{l @{\hskip 0.6cm} l @{\hskip 0.6cm} l}
		\toprule
		Parameter                          	& Value            	& Unit         			\\
		\hline\hline
		Unloaded Q            			   	& 8754 			  	&              			\\ % updated
		Operating frequency                 & 2998.46          	& MHz          			\\ % updated
		Mode separation						& 17.96				& MHz 					\\ % updated
		Structure filling time $\tau$	   	& 0.465			  	& $\mu$s 				\\ % updated
		RF power                           	& 5                	& kW           			\\ % updated
		Shunt impedance ($\beta=1$)			& 1.71				& M$\Omega$				\\ % updated
		Duty cycle                         	& 1   		     	& \%           			\\ % updated
		Repetition frequency               	& 1        			& kHz           		\\ % updated
		Average RF power                   	& 50       		   	& W            			\\ % updated
		Peak on-axis E field               	& 8.3             	& MV m$^{-1}$   		\\ % updated
		Peak surface E field               	& 25               	& MV m$^{-1}$   		\\ % updated
		Field balance 						& 1.0 : 1.0		 	& 			 			\\ % updated
		Cooling water temperature          	& 298.15          	& K            			\\ % updated
		Steady state temperature 		   	& 300.65		 	& K      				\\ % updated    
		\bottomrule
	\end{tabularx}
	\label{table:source_parameters}
\end{table}
\indent Typically, the edges of the coupling slots (`waveguide-cavity coupling' in Fig. (\ref{fig:cavity_dimensions_x_y})) are also optimized in order to minimize the maximum surface magnetic field along the coupling slot, which in turn reduces the steady-state temperature of the coupling slots' edges. This temperature is typically lowered as the blend radius closer to the cell is increased \cite{Xiao_2005}. Finally, the inner radius, outer radius, and the rounding of the magnetic coupling slots are optimized in such a way that the two cavities are coupled sufficiently and the power dissipation along the edges of the slots does not lead to excessive heating (see Fig. (\ref{fig:cavity_dimensions_x_y})).\\
\indent Optimizing the coupling slots has consequences for the separation of the frequencies of the two resonant modes. The mode-separation $f_{0}-f_{\pi}$, where for the presented structure $f_{0}$ is always the higher frequency of the two, has been maximized by variation of the sizes and shapes of the coupling slots. A mode separation of $\sim18$ MHz is obtained, with $f_{0}=2998.46$ MHz and an unloaded quality factor $Q_{0}=8754$, resulting in a 1/$e$ filling time of $\tau\approx0.47$ $\mu$s. The structure will be limited to operation at 1 kHz because of the repetition frequency of the ionization laser and the MOT loading rate. 

\subsection{Waveguides and higher-order-modes}\label{subsec:waveguides}

The design of the proposed DC/RF source features a double coupling design at the second cell. One of these couplers is connected to a WR284 waveguide with a standard length (72.136 mm) and adjusted width of 20 mm. The auxiliary port may also be used as a pump opening and can potentially house a pick-up probe \cite{Dowell_2008}.
\\
\indent The auxiliary waveguide has the same width as the feed waveguide, i.e. 20 mm (the same width as the second cell) but a smaller length (40 mm, see Fig. (\ref{fig:cavity_dimensions_x_y})), in order to ensure that the RF power can not propagate in the auxiliary waveguide (f$_{\textnormal{cut-off}}\approx3.7$ GHz). The suppression of the dipole field is not perfect, resulting in a minor dipole kick due to the power flow from the single waveguide. It however has negligible influence on the beam quality as will be shown in Sec. (\ref{sec:particle_tracking_simulations}).\\

\subsection{Electromagnetic field distribution}\label{subsec:fields}
\begin{figure}[t!]
	\includegraphics[width=1\linewidth]{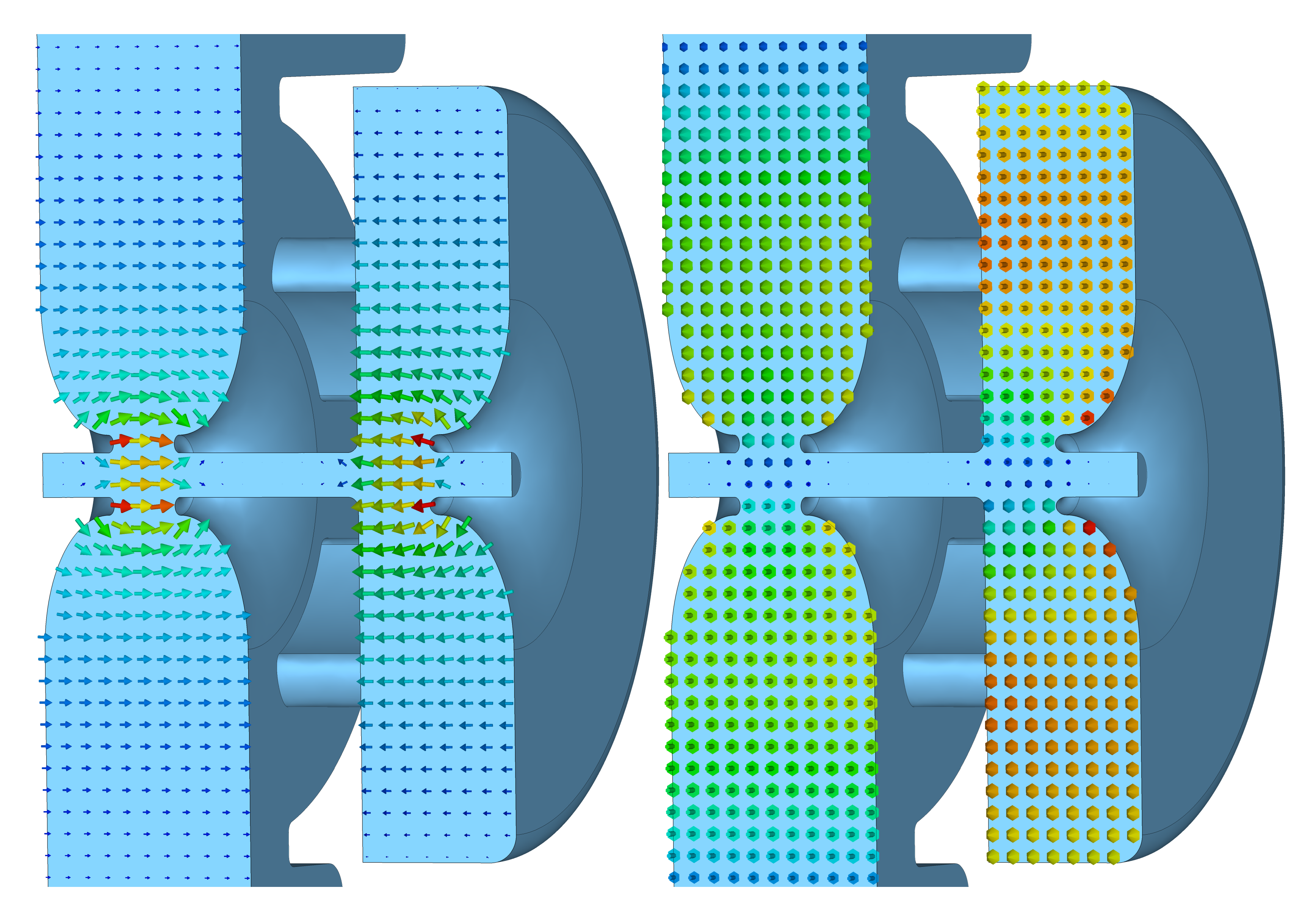}
	\caption{The electric (left) and magnetic (right) field profile where arrows indicate the direction of the field and the colors ranging from blue to red indicating increasing field strengths. The light blue structure indicates the vacuum geometry of the RF accelerator.}
	\label{fig:field_profiles}
\end{figure}

\begin{figure}[b!]
	\includegraphics[width=1\linewidth]{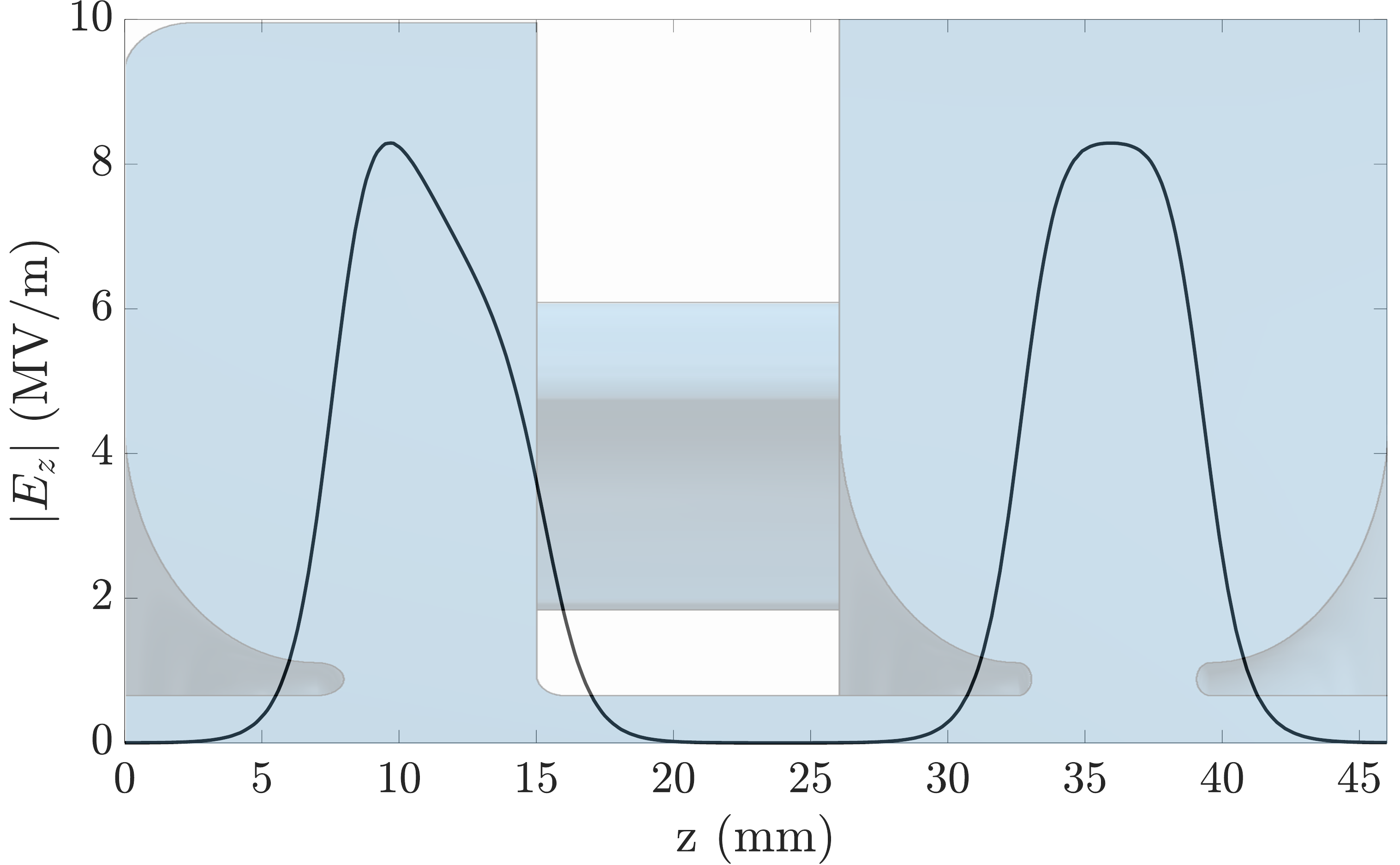}
	\caption{The on-axis electric field profile $\left|E_{z}\right|$ (black solid curve) as a function of the longitudinal coordinate $z$ for an input power of 5 kW. The RF (vacuum) structure's geometry is shown in the background (blue-gray).}
	\label{fig:on_axis_Ez_5_kW}
\end{figure}

The electromagnetic fields were calculated using the frequency-domain solver in \textsc{cst microwave studio} \cite{CST} with a simulated peak input power of 5 kW being delivered to the structure through the feed waveguide (50 W average power at a repetition frequency of 1 kHz with a 10 $\mu$s pulse length, see Table (\ref{table:source_parameters})). The resulting electric field and the transverse magnetic field $H_{x}$ are shown in Fig. (\ref{fig:field_profiles}).\\
\indent The on-axis electric field $\left|E_{z}\right|$ for a peak input power of 5 kW is shown in Fig. (\ref{fig:on_axis_Ez_5_kW}) along with an impression of the vacuum geometry of the RF accelerator. It is clearly seen that the field profile at the first cell is deformed due to the asymmetric shape of cavity I. On-axis field strengths in excess of 8 MV/m are obtained, which is sufficient for the purposes of this source. The maximum surface electric field strengths are $\sim$25 MV/m on the tip of the nose cones, well below breakdown field strengths for S-band normal-conducting copper cavities \cite{Degiovanni_2018, Vnuchenko_2020}.

\subsection{Thermal analysis}\label{subsec:thermal_analysis}
\begin{figure}[b!]
	% CST sim is found in D:\CST_files\100_keV_DC_RF_source_final_3\COUPLED_EM_THERM_VAC_and_COP_17112021_temp_scan_0p5_W
	\centering
	\includegraphics[width=0.8\linewidth]{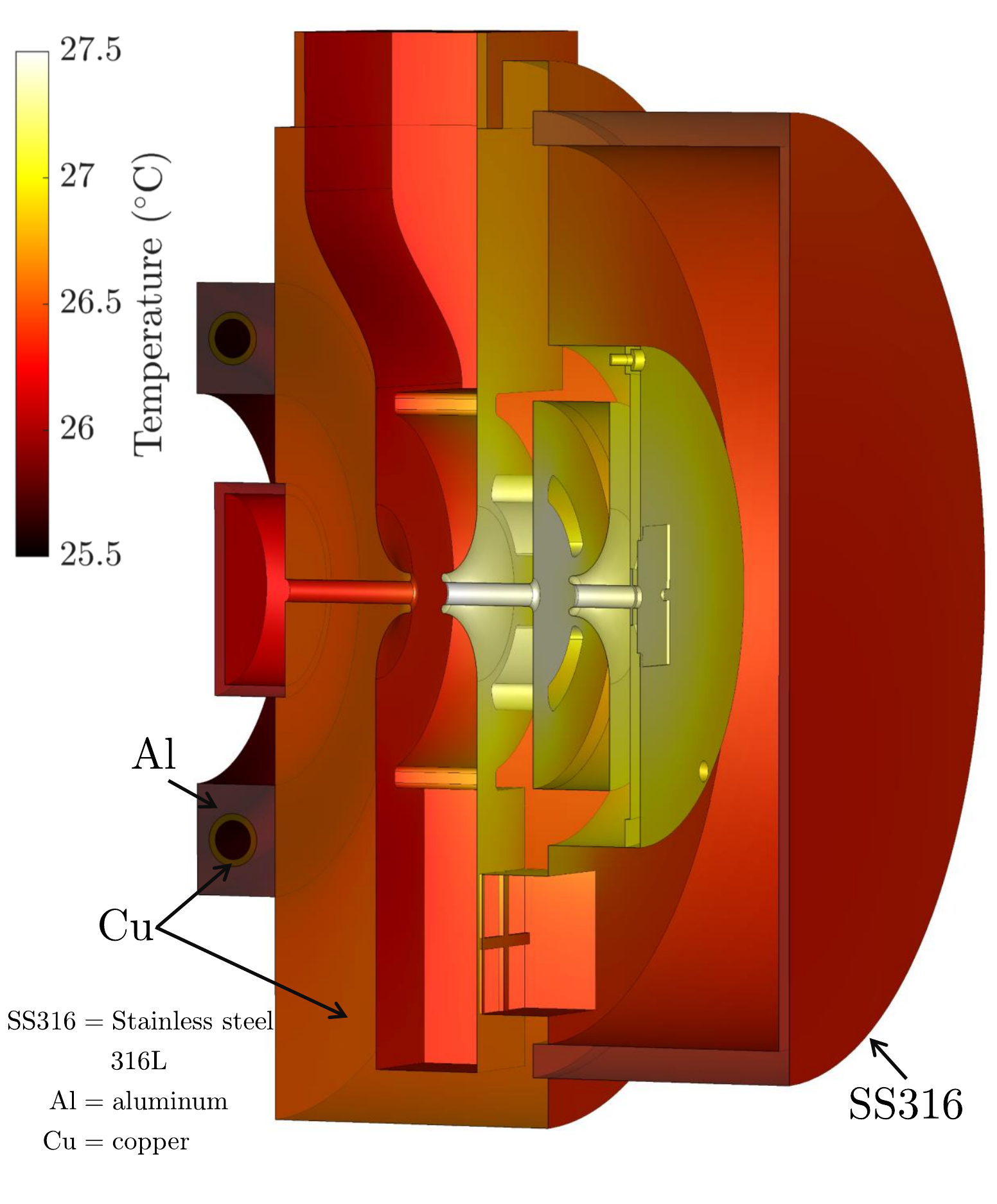}
	\caption{The steady state temperature of the accelerator for a peak input power of 5 kW, a repetition frequency of 1 kHz, and a pulse duration of 10 $\mu$s. The water cooling of the system is initialized at 25 $^{\circ}$C with a flowrate of 6 L min$^{-1}$ and a convective heat transfer coefficient of 11$\times$10$^{3}$ W m$^{-2}$ K$^{-1}$. The materials of the structure are stainless steel, copper, and aluminum (see Fig. (\ref{fig:schematic_3_4})).}
	\label{fig:thermal_simulations}
\end{figure}

During steady state operation with a peak input power of 5 kW, RF pulses with a duration of 10 $\mu$s, and repetition frequencies of 1 kHz, an average power of 50 W will be dissipated in the RF structure at the resonant frequency. The steady state heating of the RF structure is investigated using \textsc{cst microwave studio} by calculating the power dissipation at the resonant frequency and using the resulting power loss density to calculate the final temperature of the structure for some given duty cycle.\\
\indent The structure is modeled as pure copper with a thermal conductivity of 401 W K$^{-1}$m$^{-1}$ and a specific heat of 390 J K$^{-1}$kg$^{-1}$. Fig. (\ref{fig:thermal_simulations}) shows the steady state temperature of the actively cooled system with the color bar indicating the local temperature.
Fig. (\ref{fig:thermal_simulations}) shows the copper body of the RF structure along with a stainless steel CF160 pipe on the left side, which houses the accelerator module used for the static extraction field, additionally a stainless steel CF40 pipe is simulated on the right as the beginning of the beam pipe system, the cooling channel can be seen here, placed concentrically around the pipe.\\
\indent In the simulation the boundary conditions of the domain are open and the system is cooled by a hollow toroidal copper cooling channel with an inner radius of 7 mm, covered by an aluminum encasing. Water at a temperature of 25 $^{\circ}$C flows at a rate of 6 L min$^{-1}$, which results in a convective heat transfer coefficient of 11$\times$10$^{3}$ W m$^{-2}$ K$^{-1}$. In this situation the temperature change from the nose cone of the first cell to the right nose cone of the second cell is approximately +1 K. For an ideal copper pillbox cavity oscillating in TM$_{010}$ mode this would result in a frequency shift of $\frac{\partial f}{\partial T}=-51$ kHz/K at 3 GHz \cite{Pozar_1990}. This gives rise to phase shifts in the order of $\sim$3 mrad.

\section{PARTICLE TRACKING SIMULATIONS}\label{sec:particle_tracking_simulations}
\begin{figure}[b!]
	\includegraphics[width=1\linewidth]{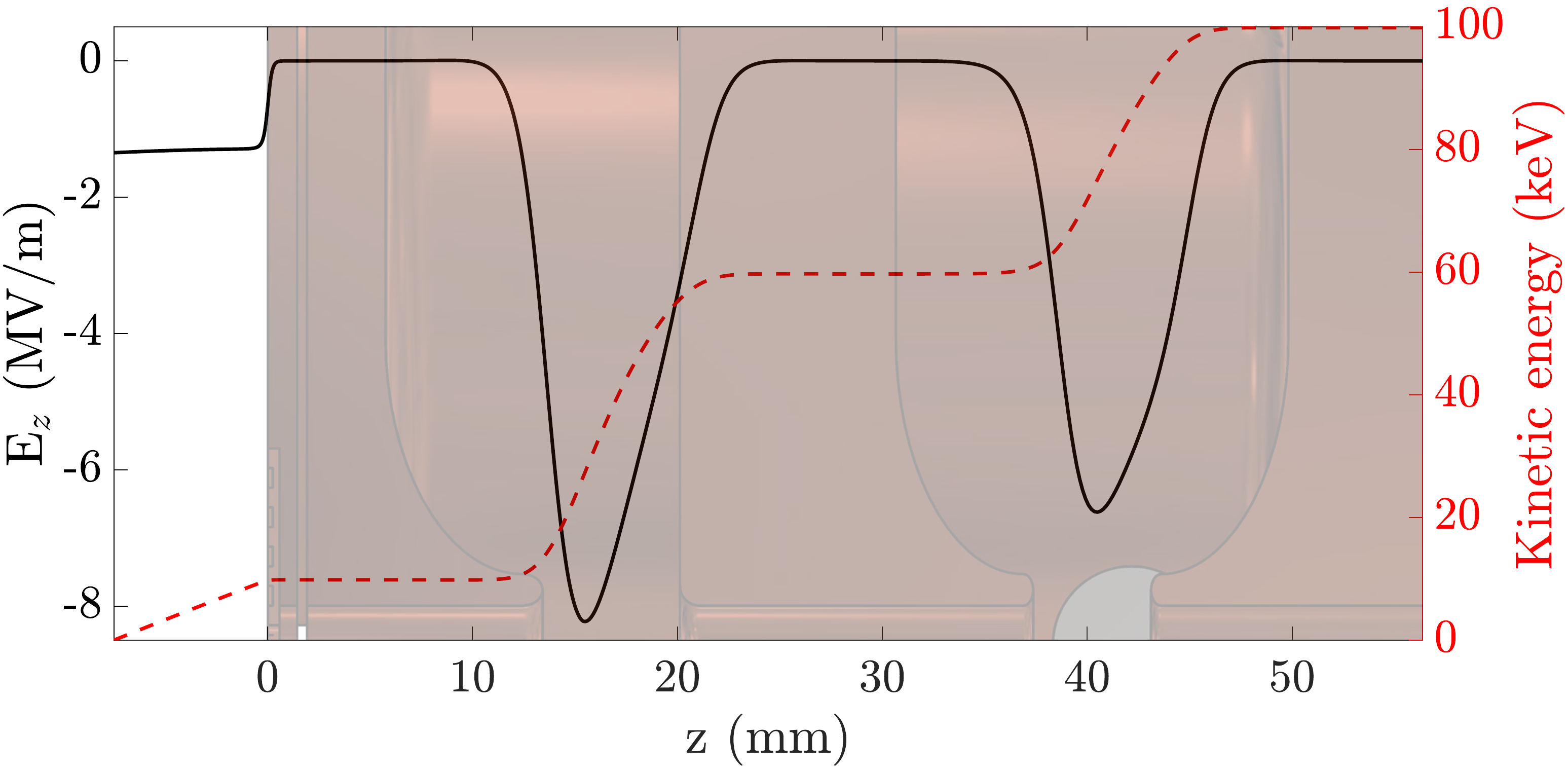}
	\caption{The average experienced electric field $E_{z}$ by the bunch (solid black) and the acquired kinetic energy (dashed red) as a function of the axial position $z$ for an input power of $P_{\textnormal{in}}\approx5$ kW. The background image is the copper housing of the RF structure.}
	\label{fig:Ez_simulation_beamline}
\end{figure}

\begin{figure}[t!]
	\includegraphics[width=1\linewidth]{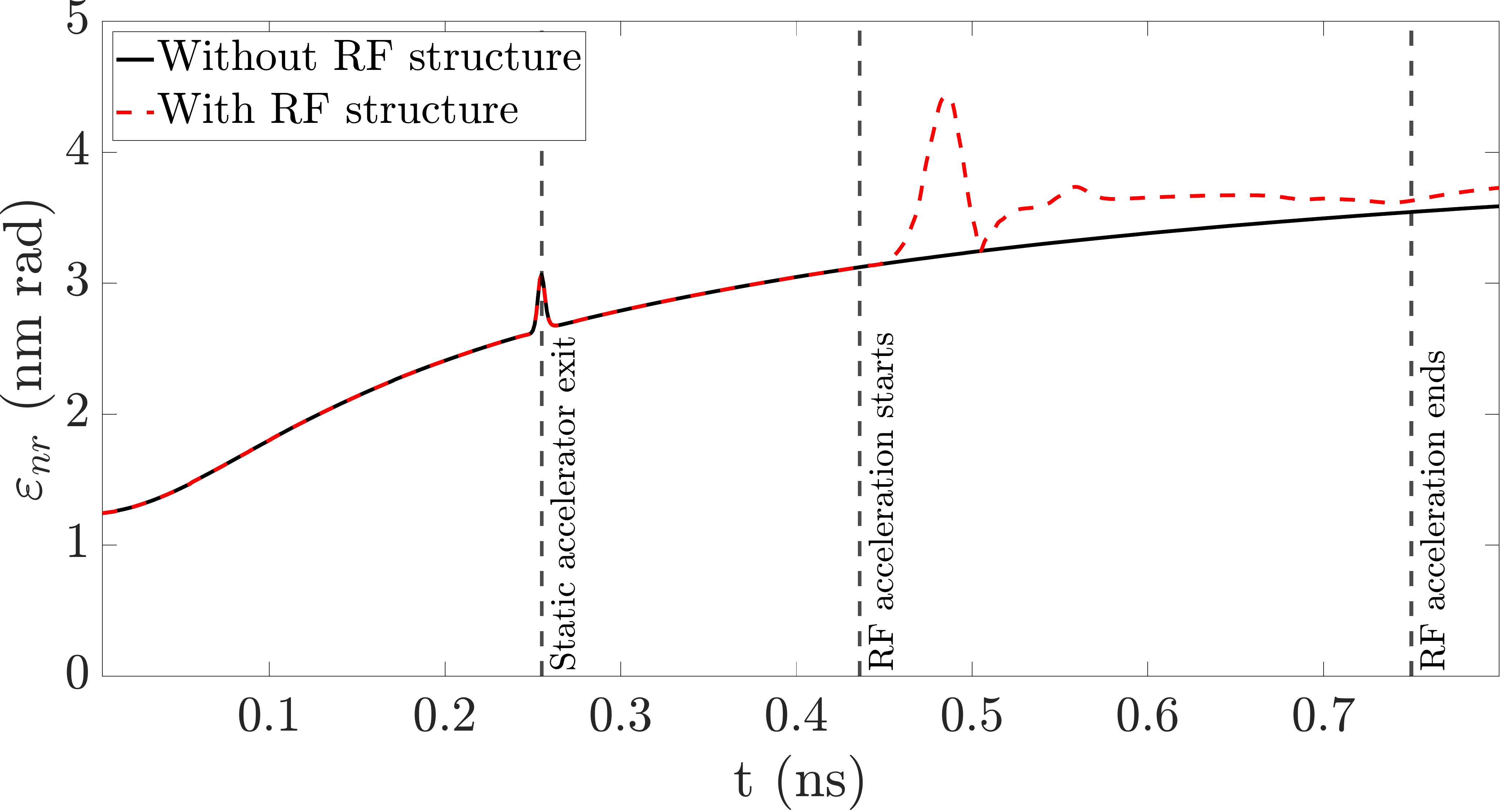}
	\caption{The normalized emittance (in the transverse direction) of a 1.6 fC bunch in a co-rotating frame where the black line signifies a simulation with the RF structure disabled and the dashed red line an identical simulation but with the RF structure enabled.}
	\label{fig:emittance_vs_sim_time_comparison}
\end{figure}

\begin{figure}[b!]
	\includegraphics[width=1\linewidth]{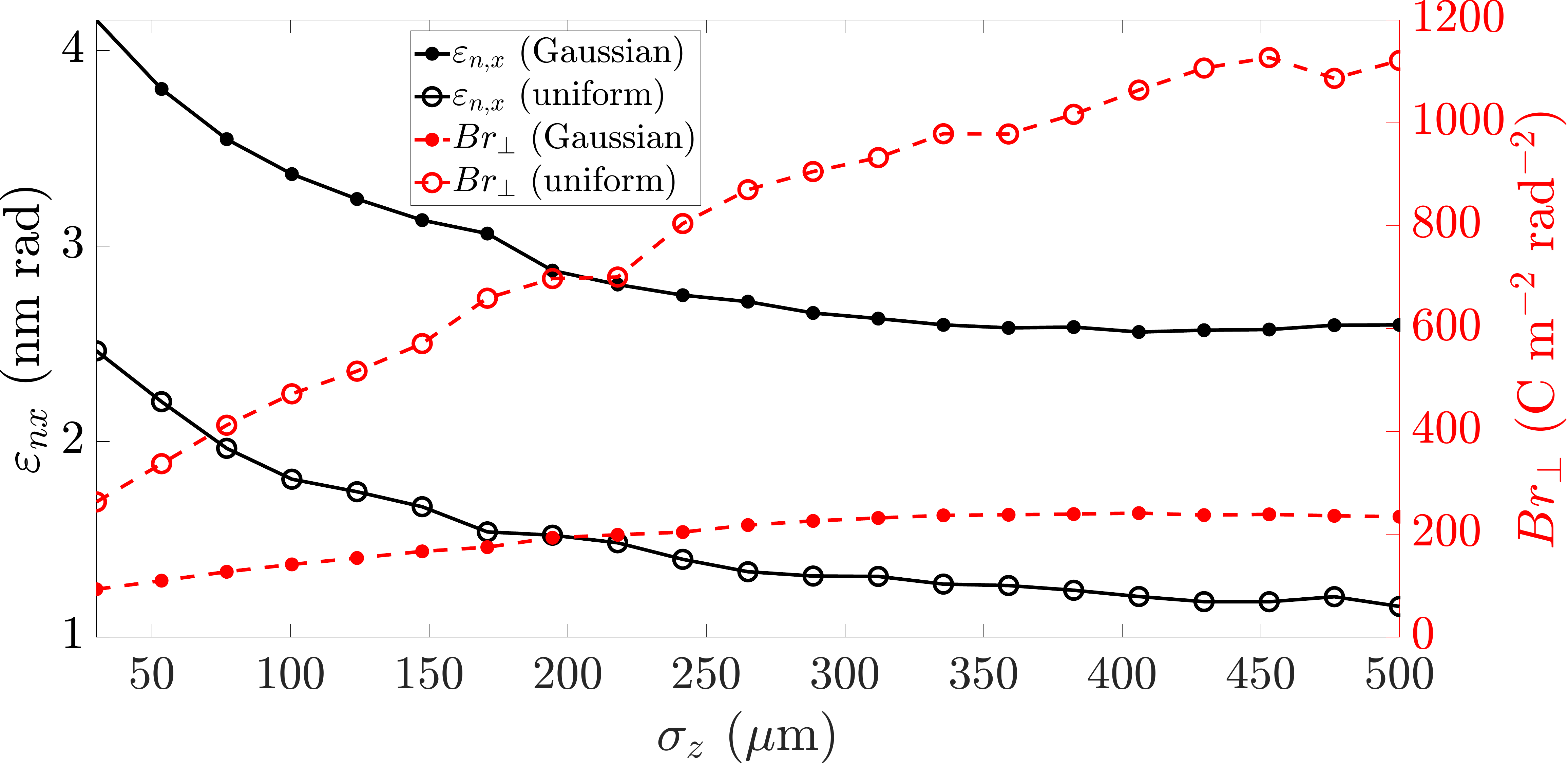}
	\caption{The normalized transverse emittance (solid black) on the left axis and the reduced transverse brightness (dashed red) on the right axis as a function of initial longitudinal rms bunch length for a Gaussian bunch (dots) and a transversely uniform bunch (circles), both with a total bunch charge of 1.6 fC.}
	\label{fig:Q_over_epsilon}
\end{figure}

\begin{figure}[t!]
	\includegraphics[width=1\linewidth]{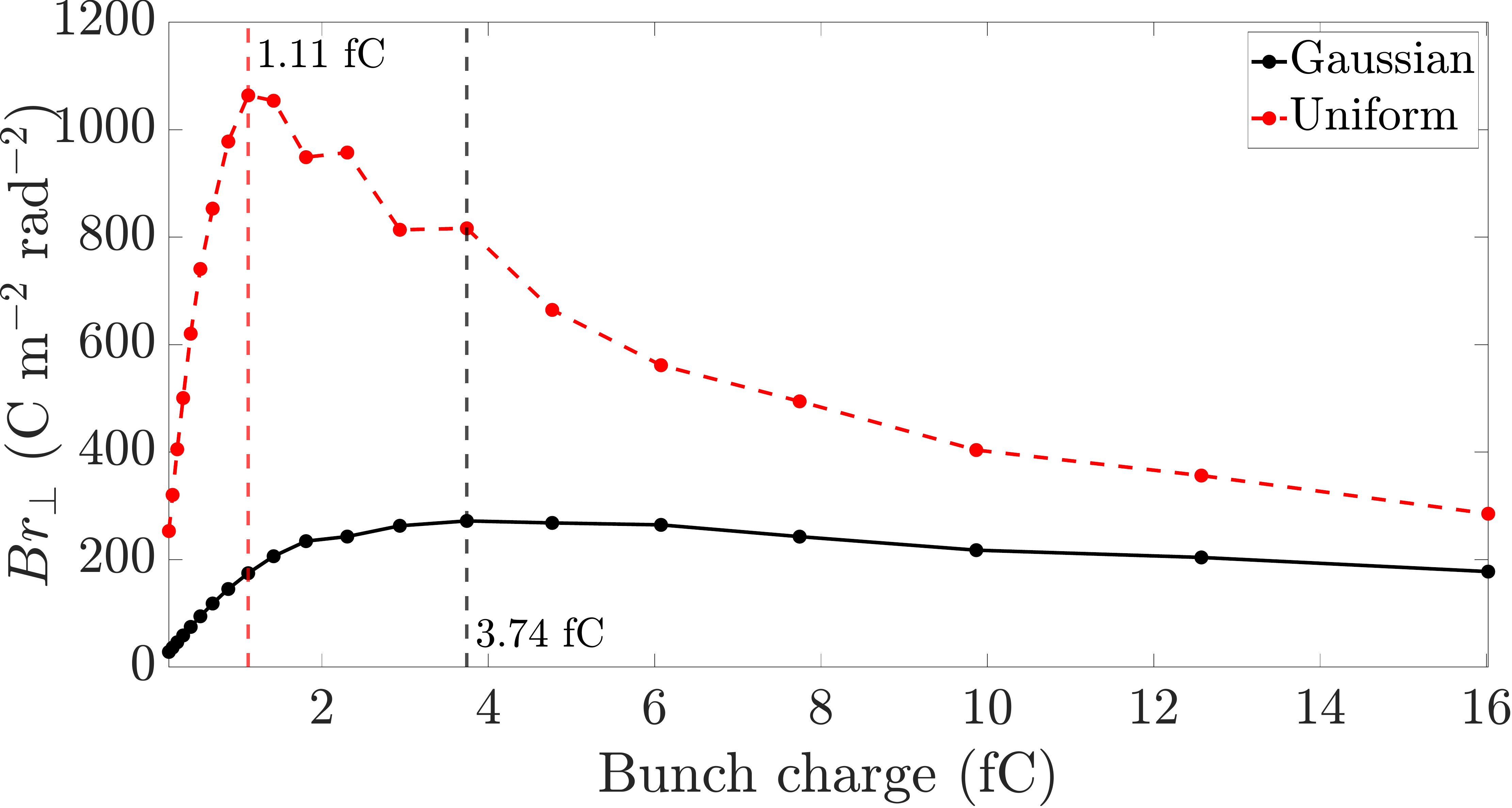}
	\caption{The reduced transverse brightness as a function of the total bunch charge for a Gaussian (solid black) and uniform (dashed red) transverse density profile.}
	\label{fig:brightness_bunch_charge}
\end{figure}

The simulations presented in this section have all been performed using the \textsc{general particle tracer} (GPT) software \cite{GPT}. A realistic beamline has been simulated which consists of a set of anti-Helmholtz configured magnetic coils (forming the MOT coils) \cite{McGilligan_2017}, a DC accelerator module \cite{Franssen_2019}, and the $1\frac{3}{4}$-cell RF structure as shown in Fig. (\ref{fig:schematic_1_2}, \ref{fig:schematic_3_4}). Detailed fieldmaps of the DC accelerator module and the RF structure are used and 1.6 fC - 16 fC electron bunches are simulated. The MOT coils used in the simulation have a total of 196 windings each and are supplied with a current of 9.0 A (front coil) and -11.3 A (back coil). Each has a radius of 90 mm and are separated by 180 mm. This creates a magnetic field gradient of $\partial B_{z}/\partial z\approx0.15$ mT/m. The DC accelerator that extracts the electron bunches generates a static field of 1.4 MV/m across a gap of 19.2 mm. The electron bunch is extracted from a $30\times30\times30$ $\mu\textnormal{m}^{3}$ (rms) ionization volume at a distance of 7.5 mm from the grating chip. The initial isotropic momentum spread is thermal with an electron temperature $T=10$ K.
\noindent The ionization process is simulated by creating electrons in the ionization volume according to a Gaussian distribution at a 1 ps rms time scale \citep{Raadt_2023}. Coulomb interactions are taken into account between all individual electrons. The interaction between the electrons and the rubidium ions is neglected in this simulation, which is justified by the fact that the electrons leave the DC field after a fraction of the inverse plasma frequency $\omega_{p}^{-1}\gg\tau_{acc}\approx0.3$ ns. This implies that the bunches are not significantly affected through disorder-induced heating \citep{Killian_2007}.\\
\indent The average electric field $E_{z}$ experienced by the electrons in the simulation and their average kinetic energy is shown in Fig. (\ref{fig:Ez_simulation_beamline}) as a function of the longitudinal position $z$.
Electrons are accelerated up to 10 keV in the static accelerator field which is located in the white region of the figure. The first and second cell of the RF structure add approximately 50 keV and 40 keV respectively. The difference in experienced field strength and corresponding acceleration in the cells is due to the finite travel time between them, resulting in a phase advance. 
\\
\indent In order to check the influence of the RF structure on the beam quality a simulation is done without the RF fields enabled in the simulation. Comparing the normalized emittance in a transverse direction as a function of the simulation time then gives an indication of the extent with which th RFe accelerator structure influences the transverse bunch dynamics, this is shown in Fig. (\ref{fig:emittance_vs_sim_time_comparison}).
Here the normalized transverse emittance in a co-rotating frame $\varepsilon_{nr}$ is shown,
which has the advantage that artificial emittance growth due to strong $x-y'$ and $y-x'$ coupling whilst traversing solenoid fields is omitted. This makes it easier to observe the actual emittance growth at the source better \cite{GPT}.
\\
\indent 
This variable is plotted as a function of simulation time for simulations with and without an RF structure, signified by the dashed red curve and the solid black curve respectively. The exit of the static accelerator is marked by the vertically dashed lines, as well as the beginning and ending of the RF acceleration. These simulations show that there is some minor growth of the beam emittance due to the RF structure: $\sim90$ pm rad at the end of the RF acceleration with respect to the non-accelerated case. This implies that the emittance growth in this simulation is driven by non-linear space charge forces, acting mainly on the periphery of the bunch.
\\
\indent The figure of merit to optimize for this source is the reduced transverse brightness, defined as:
\begin{equation}\label{eq:reduced_brightness}
	Br_{\perp}\equiv\frac{Q_{\textnormal{bunch}}}{\varepsilon_{nx}\varepsilon_{ny}},
	\end{equation}
\noindent where $Q_{\textnormal{bunch}}$ is the total bunch charge and $\varepsilon_{ny}$ the normalized transverse emittance in the $y$-direction. Bunches with an initial Gaussian distribution in the longitudinal and transverse direction are not the most optimal in terms of suppressing space charge driven emittance growth. This results in a reduced transverse brightness that does not benefit greatly from an increased bunch charge, as this exacerbates these forces. Space charge driven transverse emittance growth can be reduced by increasing the initial longitudinal length of the bunch, resulting in a larger value for the reduced transverse brightness. For 1.6 fC Gaussian bunches this is shown by the solid black and dashed red lines with dots in Fig. (\ref{fig:Q_over_epsilon}).
This figure shows the benefit of increasing the bunch length for a fully Gaussian shaped bunch with an initial rms transverse size of 30 $\mu$m, resulting in a smaller reduced transverse emittance.\\
\indent Other distributions are more suited to handle space charge driven quality degradation. Bunches with a uniform disk-like profile in the transverse plane (with a radius of 30 $\mu$m) and a Gaussian longitudinal density profile benefit much more from an increase in the initial rms longitudinal bunch length, as is shown by the solid black and dashed red lines with circles in Fig. (\ref{fig:Q_over_epsilon}). Realizing these initial distributions would however require additional optics like spatial light modulators\\
\indent In fig. (\ref{fig:brightness_bunch_charge}) the resulting reduced transverse brightness is plotted when scanning the total bunch charge up to 16 fC for transversely Gaussian and transversely uniform bunches, both  with an initial rms length of 500 $\mu$m (Gaussian). From this figure it is seen that the bunches with a uniform transverse density profile perform better, but reach their best performance at a bunch charge at around 1.1 fC with an accompanying $Br_{\perp}$ of approximately 1060 C m$^{-2}$ rad$^{-2}$, whereas the fully Gaussian bunch has its optimum at a bunch charge of roughly 3.7 fC with a corresponding $Br_{\perp}$ of roughly 272 C m$^{-2}$ rad$^{-2}$.
Additional optimization of, among other things, the initial RF phase of the accelerator structure, the field balance, initial bunch length, etc. might result in better performance. It is evident however, from Fig (\ref{fig:Q_over_epsilon}) and (\ref{fig:brightness_bunch_charge}), that a distribution which more closely resembles a homogeneously filled disk in the transverse direction will outperform a simple 3D Gaussian distribution.

\section{SUMMARY}\label{sec:summary}
The design, optimization, and particle tracking simulations of a compact 1$\frac{3}{4}$-cell, standing wave RF structure was presented. The RF structure, combined with the UCES \citep{Franssen_2019_2}, results in a hybrid DC/RF structure which was readily proposed \cite{Geer_2014}. The system presented in this article is capable of accelerating electrons up to 100 keV with minimal normalized transverse emittance growth. The structure is optimized to be resonant at approximately 2.99855 GHz and only requires 5 kW peak input power, easily supplied by commercially available solid-state amplifiers. Thermal simulations show that operation at a repetition frequency of 1 kHz can be sustained with easily manageable temperature rises.
\\
\indent The quality degradation of the electron bunches is observed primarily in the static extractor field of the source, the RF structure itself adds little to the degradation of the bunch. By using a more complex initial transverse distribution of the created bunches, the reduced transverse brightness of this source can exceed the performance of a conventional RF photosource, potentially reaching a reduced transverse brightness of  $B_{\perp}\approx10^{3}$ C m$^{-2}$ rad$^{-2}$.

\section*{Acknowledgments}

The author would like to thank A. Rajabi, T.G. Lucas, and W.F. Toonen for their support and fruitful discussions, additionally, thanks to E. Rietman, M. van der Sluis, H. van Doorn, and H. van den Heuvel for their expert technical assistance. This publication is part of the project ColdLight: From laser-cooled atoms to coherent soft X-rays (with project number 741.018.303 of the research programme ColdLight) which is (partly) financed by the Dutch Research Council (NWO).

\bibliographystyle{APSREV4-2}
\bibliography{100keVDCRF_library.bib}

\end{document}